\documentclass[british,english]{article}
\usepackage{helvet}

\usepackage[T1]{fontenc}
\usepackage[latin9]{inputenc}
\usepackage{graphicx}
\usepackage{babel}
\begin{document}
\title{The Comprehensive Blub Archive Network: Towards Design Principals
for Open Source Programming Language Repositories}
\author{Seamus Brady, seamus@corvideon.ie}
\date{28/03/2013}
\maketitle
\begin{abstract}
Many popular open source programming languages (Perl, Ruby or Python
for example) have systems for distributing packaged source code that
software developers can use when working in that particular programming
language. This paper will consider the design principals that should
be followed if designing such an open source code repository.
\end{abstract}

\section{Introduction}

Imagine that there is a software company that owns an average programming
language which we shall call Blub \cite{1} . This hypothetical company
is going bust and in true altruistic fashion they open source the
entire Blub language.

A small but enthusiastic community grows up around Blub and they decide
they would like to emulate the older, more developed open source languages.
It is decided that Blub needs it's own open source language repository
like CPAN (the Comprehensive Perl Archive Network). This is a \foreignlanguage{british}{centralised}
system that manages packaged source code (and associated metadata)
that is available for download and installation onto a developers
machine.\cite{2}. The Comprehensive Blub Archive Network (CBAN) will
be an online Blub code repository that all can share and use. 

However some worries are expressed after availability and security
issues were reported with the Ruby gems website \cite{3} (the Ruby
language analogue of CPAN - in Ruby pre-packaged code is known as
a ``gem'' \cite{4}). It is decided that a set of design principals
for a secure and reliable open source code repository such as CBAN
will be drawn up to mitigate any future problems.

This paper addresses this hypothetical challenge.

In Section 2, the reliability and security challenges facing an open
source code repository are outlined. An set of solutions is identified
to address these challenges. In Section 3, the six design principals
arising out of these solutions are discussed. Related work is discussed
in Section 4.

\subsection{Resources and Methods}

This problem was approached by researching existing open source code
repositories for Perl, Ruby and Python and examining their strengths
and weaknesses. I also looked at operating systems package managers
such as Debian's apt-get.

It should be noted that while there is some criticism of existing
open source code repositories in this paper, no existing system embodies
all the design principals for an open source code repository. There
is much to be done in all systems, even in the most developed system
CPAN, especially on mirror and package signatures.

\section{Background}

\subsection{The Challenges: Reliability and Security }

If CBAN were to adopt a Prime Directive it would be ``Trustworthy
Software'' as identified by John Chambers \cite{5}
\begin{quote}
\textquotedbl ...the software provider {[}has{]} a strong responsibility
to produce a result that is trustworthy, and, if possible, one that
can be shown to be trustworthy...\textquotedbl{}
\end{quote}
A brief analysis of the rubygems.org \cite{3} hacking incident reveals
two different challenges:
\begin{itemize}
\item When the rubygems.org system administrators noticed that the server
had been hacked, the system was taken offline. This meant that developers
trying to install gems (packaged Ruby code) were unable to do so.
The system had no reliability built in for incidents such as this.
This was a single point of failure.
\item As the rubygems.org system was compromised, all gems on the system
were no longer trustworthy. A full security audit had to be done before
the system could go back online. The gems installed on developers
machines just before the rubygems.org server went offline could no
longer be considered ``safe''.
\end{itemize}
The full impact of compromised gems (used in some popular web development
stacks such as Ruby on Rails) was outlined by Patrick McKenzie, a
Ruby developer:
\begin{quote}
One of my friends who is an actual security researcher has deleted
all of his accounts on Internet services which he knows to use Ruby
on Rails. That\textquoteright s not an insane measure. \cite{6}
\end{quote}
So when an open source code repository has problems, these problems
tend to break down across two axis:
\begin{itemize}
\item \textbf{Reliability}: when an open source code repository goes offline,
developers may not be able to work due to lack of code availability.
\item \textbf{Security}: Compromised packaged code may make it onto multiple
computers before any intrusion is even detected. Developers must be
able to trust and identify secure code packages.
\end{itemize}
If we wish to fulfill CBAN's Prime Directive of ``Trustworthy Software'',
CBAN has to be \textbf{reliable} and \textbf{secure}.

\subsection{Solutions to Challenges}

There are three main solutions to this challenge - mirrors, signatures
and standards. 

\subsubsection{Mirrors}

Reliability in a distributed web system can have many different facets
- availability, performance, cost \cite{7}. Let us assume that all
other system operations such as backups, load balancing etc. are taken
care of. What would be the most important remaining factor in avoiding
downtime for developers using CBAN? 

The simplest answer is built in redundancy:
\begin{quote}
If there is a core piece of functionality for an application, ensuring
that multiple copies or versions are running simultaneously can secure
against the failure of a single node.

Creating redundancy in a system can remove single points of failure
and provide a backup or spare functionality if needed in a crisis.
\cite{7}
\end{quote}
This is done by providing CBAN with a set of mirrors. A duplication
of the packaged source code and it's associated metadata will be available
across several official public servers and replicated from a master
server. For instance, CPAN has it's own global set of mirror servers
\cite{8}. The improvement of the mirroring system for Ruby gems is
under discussion \cite{9}.

The full system for CBAN would also involve several other auxiliary
servers for code searches, source code control and other administrative
functions as displayed in Figure 1.

\begin{figure}
\centering{}\includegraphics[scale=0.5]{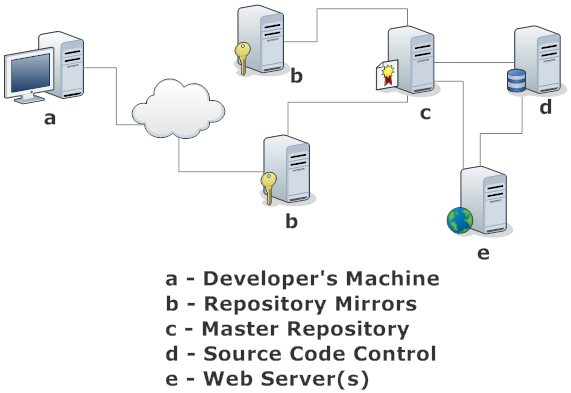}\caption{Repository and Associated Servers}
\end{figure}

\subsubsection{Signatures}

When code becomes compromised on a public code repository, an attack
on a user of the repository can come from several different directions
\cite{10}:
\begin{itemize}
\item The system may download and install arbitrary packages containing
a destructive payload.
\item The system may download and install older or out of date packages
with know security holes that can be exploited.
\item The system may stop the download and installation of newer packages
that fix security holes.
\item The system may download extra ``unsafe'' packages by masking them
as dependencies to known ``safe'' packages.
\end{itemize}
\begin{figure}
\centering{}\includegraphics[scale=0.5]{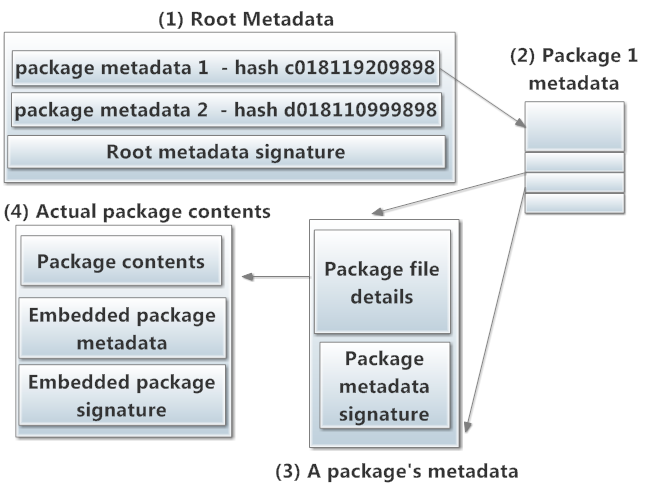}\caption{Repository Internal Structure (based on Cappos, Samuel, Baker and
Hartman\cite{10}) }
\end{figure}

These attacks can be avoided by using signing on both the code package
and the metadata associated with the code. Figure 2 gives an overview
of the structure of a code repository.
\begin{itemize}
\item At (1) is the root metadata - this is a list of compressed package
files, the packages secure hashes and the root metadata signature.
This signature stops tampering of the contents of each package, and
if associated with a timestamp, allows a way to check if a mirror
has an up-to-date and correct copy of the root metadata. 
\item Each code package has metadata as listed at (2) where we can see the
package contents and package metadata signature. This can be matched
against the embedded package metadata in the package file (3) itself
to check that the package is safe. 
\end{itemize}
Enforcing both these levels of signature checking will stop the attacks
listed above as the data can be checked at the various stages of deployment:
\begin{itemize}
\item Root metadata can be used to secure mirror-to-mirror replication including
an enforced timeout for updates.
\item Package level signatures can be used to secure dependency checking
and installation.
\end{itemize}
Again, if the rubygems.org system enforced signatures even at the
packet level, no compromised code would have been installable from
the system. This has been noted as an inadequacy and is currently
being addressed \cite{11}.

\subsubsection{Standards}

This is a more nebulous solution than then previous two, but important
nonetheless. Standards are important as they allow predictable results.
Predictable results means better tooling and automation. Better tooling
and automation lead to more easily scaled systems. Standards do not
need to be set in stone of course, they evolve with the language.
Nevertheless, if simple standards are adopted early on, it will allow
CBAN to grow and evolve gracefully.

\paragraph{Packaging and deployment:}

This need for adoption of standards can be seen in the problems and
challenges facing a Python developer when packaging their code for
deployment \cite{12}. 
\begin{quote}
The absence of good packaging standards has complicated the development
of Python package managers. Installers have trouble finding the most
recent version of packages due to the lack of a complete standard
for package version numbers. Pip must create a nonstandard record
of files installed for each package to support uninstallation. Setuptools
had to introduce its own method of defining dependencies to work with
the way packages are distributed. These nonstandard additions have
helped address real problems, but have contributed to the fragmentation
of the Python packaging ecosystem. \cite{13}
\end{quote}
If packaging and deployment standards are adopted, then the CBAN system
can develop the way that the Perl CPAN system has developed. As the
Test Anything Protocol has been adopted by Perl as a standard, every
module downloaded from CPAN can provide it's own test suite that is
run on installation \cite{15}. Building on such standards, Perl authors
can now use such tools as Dist::Zilla to automate their CPAN deployments
\cite{16} and run tests across multiple operating systems automatically
by using CPAN Testers \cite{17}.

\paragraph{Licensing and code ownership policy:}

Similar problems can happen when the licensing is not \foreignlanguage{british}{standardised}
across a project, holding back contributions while legal advice goes
back and forth. See the long drawn out history of the Squeak programming
language and the Apple license for instance \cite{14}.

Policies regarding code ownership should also be standardised. It
may be useful to distinguish code ownership from code authorship in
some cases. For instance, can ownership of code revert back to the
community if the original author abandons it? Can the community hand
the ownership over to another maintainer? These issues can be problematic
in an open source community when people cannot continue their commitment
to projects they started or someone else tries to take over from an
existing author \cite{18}. 

Also, as the signature section above outlined - knowing who wrote
and signed the code package you want to install is important information.

\paragraph{Namespace conflicts}

The programming language should have some method for allowing similarly
named modules to exist in the system in different packages without
causing conflict. 

Adopting a standard on this early in the process of building CBAN
will allow greater flexibility for would-be code contributors who
won't have to worry about managing module naming conflicts in a global
namespace.

\paragraph{Other areas}

There are many other areas where standards can be adopted such as
community policies on spam, governance and grant aid for developers.
Any standards here would help CBAN develop and flourish. These are
implied here, rather than listed exhaustively.

\section{Results - The Six Design Principals}

The six design principals below can be abstracted away from the discussion
above. Each of these design principals has one or more statements
that can be used as a standard for developing a system such as CBAN.
\begin{itemize}
\item Distributed Systems Design 

\begin{itemize}
\item The system should avoid a single point of failure by providing a set
of public, secure mirrors.
\item Mirrors should be authenticated against the master repository using
signed root metadata and an expiration timestamp. 
\end{itemize}
\item Package Reliability 

\begin{itemize}
\item Namespaces should be provided so that naming conflicts do not occur.
\item Correct metadata should be provided for dependency management. 
\item A test suite should be provided by the author that can be run on deployment.
\item Reviews and reports on packages should be available on an associated
web site.
\end{itemize}
\item Package Security 

\begin{itemize}
\item Package level signing should be available so that compromised packages
can be identified.
\item Changelogs and history for each package should be available for public
audit.
\end{itemize}
\item Source Management 

\begin{itemize}
\item All contributed code should be licensed on one or more standard open
source licenses to avoid legal issues.
\item Library authorship should be distinct from library ownership so that
abandoned projects can be managed.
\item Source code control should be put in place.
\end{itemize}
\item Reliable Deployment 

\begin{itemize}
\item The system should provide standard packaging and deployment tools
to encourage automation.
\end{itemize}
\item System Manageability 

\begin{itemize}
\item Standard documented policies should be developed to encourage procedural
openness, commitment and involvement.
\item The system should be managed by community via an open and accountable
organisation - as the Zen of CPAN says...

\begin{quote}
Perhaps the most demanding thing is commitment: someone must keep
things running. A slowly decaying and dusty archive is almost worse
(and certainly more sad) than no archive at all \cite{19}.
\end{quote}
\end{itemize}
\end{itemize}

\section{Related Work }

No work has been done directly on the design principals for open source
language repositories. Several short histories of existing open source
language repositories have been written however \cite{12,19}.

Research work has been done on related systems such as operating system
package management. Professor Justin Cappos of Polytechnic Institute
of NYU has published in this area \cite{20}.

\section{Summary }

Six design principals for designing an open source code repository
have been outlined.

These are:
\begin{itemize}
\item Distributed Systems Design 
\item Package Reliability 
\item Package Security 
\item Source Management 
\item Reliable Deployment 
\item System Manageability 
\end{itemize}


\begin{thebibliography}{10}
\bibitem[1]{1}Beating the Averages. http://www.paulgraham.com/avg.htmlhttp

\bibitem[2]{2}http://www.cpan.org/misc/cpan-faq.html\#What\_is\_CPAN

\bibitem[3]{3}RubyGems.org hacked, interrupting Heroku services and
putting sites using Rails at risk. http://bit.ly/WA4o7n

\bibitem[4]{4}http://rubygems.org/pages/about

\bibitem[5]{5}Page 4, Chambers, John M. (2008). Software for Data
Analysis: Programming with R. Springer. ISBN 0-387-75935-2.

\bibitem[6]{6}What the Rails security issue means for your startup.
http://www.kalzumeus.com/2013/01/31/what-the-rails-security-issue-means-for-your-startup/

\bibitem[7]{7}Scalable Web Architecture and Distributed Systems.
http://www.aosabook.org/en/distsys.html

\bibitem[8]{8}http://mirrors.cpan.org/

\bibitem[9]{9}https://github.com/rubygems/rubygems-mirror/wiki/Mirroring-2.0

\bibitem[10]{10}J. Cappos, J. Samuel, S. Baker and J. Hartman. A
Look In the Mirror: Attacks on Package Managers, 2008. http://isis.poly.edu/\textasciitilde jcappos/papers/cappos\_mirror\_ccs\_08.pdf

\bibitem[11]{11}A Practical Guide to Using Signed Ruby Gems. http://blog.meldium.com/home/2013/3/3/signed-rubygems-part

\bibitem[12]{12}Python Packaging. http://www.aosabook.org/en/packaging.html

\bibitem[13]{13}The Future of Python Packaging. https://us.pycon.org/2012/schedule/presentation/498/

\bibitem[14]{14}http://squeak.org/SqueakLicense/?version=4

\bibitem[15]{15}http://search.cpan.org/\textasciitilde petdance/Test-Harness-2.64/lib/Test/Harness/TAP.pod

\bibitem[16]{16}Dist::Zilla.http://dzil.org/

\bibitem[17]{17}http://static.cpantesters.org/page/about.html

\bibitem[18]{18}http://help.rubygems.org/discussions/questions/55-someone-has-acquired-ownership-of-a-gem-without-my-permission

\bibitem[19]{19}http://www.cpan.org/misc/ZCAN.html

\bibitem[20]{20}http://isis.poly.edu/\textasciitilde jcappos/publications.html
\end{thebibliography}
\end{document}